\documentstyle[axodraw,colordvi]{frank}

\newenvironment{comment}[1]{}{}
\def\beq{\begin{equation}}
\def\eeq{\end{equation}}
\def\bea{\begin{eqnarray}}
\def\eea{\end{eqnarray}}
\def\barr{\begin{array}}
\def\earr{\end{array}}
\newcommand{\sm}{standard model}
\newcommand{\cm}{center of mass}
\newcommand{\xs}{cross section}
\newcommand{\lc}{linear collider}
\newcommand{\pe}{\mbox{$e^+e^-$}}
\newcommand{\ee}{\mbox{$e^-e^-$}}
\newcommand{\ep}{\mbox{$e^-\gamma$}}
\newcommand{\pp}{\mbox{$\gamma\gamma$}}
\newcommand{\sw}{\mbox{$\sin\theta_w$}}
\newcommand{\cw}{\mbox{$\cos\theta_w$}}
\newcommand{\swt}{\mbox{$\sin^2\theta_w$}}

\newcommand{\swf}{\mbox{$\sin^4\theta_w$}}
\newcommand{\cwf}{\mbox{$\cos^4\theta_w$}}

\begin{document}

\begin{flushright}
PSI-PR-96-16\\
MPI-PhT/96-49\\
June 1996
\end{flushright}

\vfill

\begin{frontmatter}
\title{Polarization and the Weak Mixing Angle \\
        in High Energy $e^\pm e^-$ Collisions}
\author{Frank Cuypers}
\address{{\tt cuypers@pss058.psi.ch}\\
        Paul Scherrer Institute,
        CH-5232 Villigen PSI,
        Switzerland}
\author{Paolo Gambino}
\address{{\tt gambino@mppmu.mpg.de}\\
        Max-Planck-Institut f\"ur Physik,
        Werner-Heisenberg-Institut,
        F\"ohringer Ring 6, 
        D--80805 M\"unchen, 
        Germany}
\begin{abstract}
At a linear collider of the next generation
the large event rates expected from Bhabha and M\o ller scattering
may be used to determine 
simultaneously
$\sin^2\theta_w$ and the polarization of both beams
with very high accuracy.
These measurements can be performed 
in parallel to the other tasks of the linear collider
as a free by-product.
A high degree of polarization 
and a good polar angle coverage of the detectors
turn out to be major assets.
\end{abstract}
\end{frontmatter}

\vfill
\clearpage

While the LHC offers an entry 
into the the high energy regime of the \sm\
with a significant opportunity for discovering new phenomena,
the linear electron colliders of the next generation \cite{lc}
will provide a complementary program of experiments
with unique opportunities 
for both discoveries and precision measurements.
A major asset to fulfill this purpose
is the versatility of the \lc s,
as they can be operated in the four \pe, \ee, \ep\ and \pp\ modes,
with highly polarized electron and photon beams.
Moreover,
starting from a \cm\ energy of several hundred GeV,
it will later be possible to upgrade these machines
into the TeV range.

An important feature of the \lc s
is the high degree of polarization
which can be obtained for the electron beams.
Beam polarizations exceeding 80\%\
are by now routinely obtained at SLAC
and are steadily improving. 
A final 90\%\ electron polarization seems a 
quite sensible assumption \cite{snowmass}.
Concerning the positron beam, 
although at present no scheme for polarizing  positrons
has  been proven to be implementable, 
there are reasonable hopes 
that some practicable technology may be available
by the time a \lc\ is operating.
This ingredient is an important additional lever arm
to increase the sensitivity of the searches for new phenomena, and the 
precision measurement which are the main goal of a \lc.
It is therefore of utter importance 
to be able to  measure the degree of polarization
with great accuracy.

We propose here a simple method 
to determine the polarization of both beams
in \pe\ and \ee\ collisions \cite{e-e-}.
This procedure takes advantage of the large \xs s
of Bhabha and M\o ller scattering
to obtain a good analyzing power,
competitive with Compton polarimetry \cite{slcpol}.
Moreover,
as the polarizations are measured
from the distributions of the final state electrons and positrons, 
we are guaranteed to 
take into account all depolarizing effects 
which can spoil the initial beam polarization
at the interaction point. 
A similar procedure has been
illustrated for the $Z^0$ peak in Ref.~\cite{flott}.

The interesting feature of this measurement
is that it simultaneously provides 
a very accurate determination of \swt.
At present, 
parity violating asymmetry measurements in $Z^0$ decays
have allowed its most precise determination:
combining the SLD measurement
of the left-right asymmetries with the 
various asymmetries from LEP, 
the effective leptonic \swt\ 
is now constrained to $0.2314\pm0.0003$ \cite{lep}.
An early discussion of the determination of the weak mixing 
angle from Bhabha scattering at LEP1 can be found in \cite{lep1}. 
After the end of operation of the \pe\ colliders on the $Z^0$ peak, 
the situation is unlikely to improve significantly, although interesting 
proposals  have been put forward, 
for both low \cite{czarnecki} and at high energy \cite{fisher} experiments.
It is therefore  particularly interesting to study the potential of a 
high energy \lc\ in this respect. 

The typical \lc\ designs
aim at an integrated yearly \pe\ luminosity $\cal L$
scaling with the squared \cm\ energy $s$
like

\beq
\label{lumpe}
{\cal L}_{e^+e^-} \mbox{ [fb$^{-1}$] } \approx 80 \,s \mbox{ [TeV$^2$] } ,
\eeq
or $  {\cal L}_{e^+e^-} \approx 3 \times 10^7 s $ in $c=\hbar=1$ units.
For the luminosity of the \ee\ mode 
we take

\beq
\label{lumee}
{\cal L}_{e^-e^-} \approx {1\over2} {\cal L}_{e^+e^-}
~,
\eeq

because this mode will suffer to some extent from the 
anti-pinch effect \cite{jim}.
If needed,
it is straightforward to modify our results for a different scaling relation.

Using $N^+$ and $N^-$ for the number of particles in a beam which are
longitudinally polarized parallel or antiparallel to their momentum,
we define the polarization of an electron or positron beam to be

\beq
P_{e^-} = \frac{N^+ - N^-}{N^+ + N^-} 
\qquad
P_{e^+} = \frac{N^- - N^+}{N^+ + N^-}
~.
\eeq

With this definition 
$P=+\langle h \rangle$  
for  electrons and
$P=-\langle h \rangle$  
for  positrons,
where $\langle h \rangle$ is the helicity mean value.

We assume the integrated luminosities 
to be  equally distributed
over the four possible combinations of beam polarizations
$LL$, $RR$, $LR$ and $RL$ 
(with $R$ and $L$ referring to positive
and negative polarizations of the beams, respectively).   
\,   From the physics point of view
there is no difference between the last two combinations
in the \ee\ mode.
However, since the electron guns may have different efficiencies,
it is important to consider them both 
in order to measure this hardware asymmetry. 
It is essential 
that the polarization of the beams be flipped randomly
at short time intervals, 
a technique in use at SLC \cite{SLC}.
In this case,
if the absolute value of the polarization is on average constant, 
random and systematic fluctuations cancel out.

Neglecting the $Z^0$ width,
the polarized differential Bhabha and M\o ller scattering \xs s are
\bea
\label{pedxs}
&&
{d\sigma^{e^+e^-} \over dt}
=
{4\pi\alpha^2 \over s^2}\times
\\\nonumber&&\hskip-2em
\left\{
  {1 + P_1 + P_2 + P_1P_2 \over 4}
  \left[
    \left( \sum_i R_i^2 \left( {u \over s-m_i^2} + {u \over t-m_i^2} \right)
        \right)^2
    + \left( \sum_i L_iR_i {t \over s-m_i^2} \right)^2
  \right]
\right.
\\\nonumber&&\hskip-2em
+
  {1 - P_1 - P_2 + P_1P_2 \over 4}
  \left[
    \left( \sum_i L_i^2 \left( {u \over s-m_i^2} + {u \over t-m_i^2} \right)
               \right)^2
    + \left( \sum_i L_iR_i {t \over s-m_i^2} \right)^2
  \right]
\\\nonumber&&\hskip-2em
\left.
+
  {1 - P_1P_2 \over 2}
  \left( \sum_i L_iR_i {s \over t-m_i^2} \right)^2
\right\},
\eea

\bea
\label{eedxs}
&&
{d\sigma^{e^-e^-} \over dt} 
=
{2\pi\alpha^2 \over s^2}\times
\\\nonumber&&\hskip-2em
\left\{
  {1 + P_1 + P_2 + P_1P_2 \over 4}
  \left( \sum_i R_i^2 \left( {s \over t-m_i^2} + {s \over u-m_i^2} \right)
        \right)^2
\right.
\\\nonumber&&\hskip-2em
+
  {1 - P_1 - P_2 + P_1P_2 \over 4}
    \left( \sum_i L_i^2 \left( {s \over t-m_i^2} + {s \over u-m_i^2} \right)
         \right)^2
\\\nonumber&&\hskip-2em
\left.
+
  {1 - P_1P_2 \over 2}
  \left[
    \left( \sum_i L_iR_i {t \over u-m_i^2} \right)^2
  + \left( \sum_i L_iR_i {u \over t-m_i^2} \right)^2
  \right]
\right\}
~,
\eea
where 
$\alpha$ is the fine structure constant,
$P_1$ stands for the positron polarization
in the case of Bhabha scattering,
$s,\ t,\ u$ are the Mandelstam variables,  
the summations are over $i=\gamma,Z^0$ 
and the couplings are defined by
\beq
R_\gamma = L_\gamma = 1~,
\qquad
R_Z = -{\sw \over \cw}~,
\qquad
L_Z = {1-2\,\swt \over 2\,\sw\cw}~.
\eeq

It is our purpose
to use these \xs s
to determine as precisely as possible
the values of the weak mixing angle
and the polarization of each beam:
\beq
\label{param}
\swt \qquad P_1 \qquad P_2
~.
\eeq
The most natural choice of observables
for determining these parameters
are the differential events rates 
\beq 
\delta n_{LL} \qquad \delta n_{RR} \qquad \delta n_{LR} \qquad \delta n_{RL} 
~,
\eeq
where for each bin
\beq
\delta n = {\cal L} \int_{\rm bin} d\cos\theta {d\sigma \over d\cos\theta}
~.
\eeq
However,
the experimental determination
of the absolute \xs s
is hindered by the systematic error 
on luminosities, acceptances and efficiencies,
which dominate the statistical errors
when the event rates are as large as in Bhabha and M\o ller scattering.
It is therefore of great advantage 
to use three independent differential polarization asymmetries, 
for example
\bea
\label{as1}
A_1 = { \delta n_{LL} - \delta n_{RR} \over \delta n_{LL} + \delta n_{RR} }
\\\nonumber\\
\label{as2}
A_2 = { \delta n_{RR} - \delta n_{LR} \over \delta n_{RR} + \delta n_{LR} }
\\\nonumber\\
\label{as3}
A_3 = { \delta n_{LR} - \delta n_{RL} \over \delta n_{LR} + \delta n_{RL} },
\eea
for which the systematic errors cancel out
to a very large extent.
As long as the correlations between the three asymmetries
are correctly taken into account
and the statistical errors dominate, 
it does not matter which triplet of independent asymmetries is chosen.
Any choice other than (\ref{as1}--\ref{as3})
yields the same results.

We have chosen to normalize the $Z^0$ couplings by the fine structure
constant $\alpha$. 
In this way 
the asymmetries depend solely on $\sw$ and the beam polarizations,
which effectively parametrize all the available information. 
Moreover,
when it will come to compute the radiative corrections,
in the framework
of an $\overline{MS}$ scheme \cite{msbar}
this choice has the additional advantage 
of avoiding large electroweak corrections,
such as $m_t^2$ corrections.

We have checked 
that the accuracy with which the parameters (\ref{param}) can be measured
is such that
we can safely assume a linear dependence of the \xs s 
(\ref{pedxs},\ref{eedxs})
in the region of interest,
{\em i.e.},
within the error bands around the central values.
The error bands 
corresponding to one standard deviation
are therefore given by the quadratic form

\beq
\label{egg}
\left(
~\Delta\swt~ 
~\Delta P_1~ 
~\Delta P_2~ 
\right)
\quad W^{-1} \quad
\left(
\barr{c} 
\Delta\swt \\ 
\Delta P_1 \\
\Delta P_2
\earr
\right)
~=~1
~,
\eeq

where the inverse covariance matrix $W^{-1}$ is given by

\beq
\label{cov}
W^{-1}_{ij} 
=
\sum_{k,l=1}^3
\sum_{\rm bins}
V^{-1}_{kl}
\left( {\partial A_k \over \partial \epsilon_i} \right)
\left( {\partial A_l \over \partial \epsilon_j} \right)
\eeq

\beq
\epsilon_i = \swt , P_1 , P_2
~.
\eeq

In contrast to the polarized \xs s,
the asymmetries are correlated.
Their covariance matrix $V$ 
contains therefore off-diagonal terms
and is given by

\bea
\label{V}
V_{kl}
&=&
\langle (A_k-\bar A_k) (A_l-\bar A_l) \rangle
\\\nonumber
&=&
\sum_{i=1}^4
(\Delta n_i)^2
\left( {\partial A_k \over \partial n_i} \right)
\left( {\partial A_l \over \partial n_i} \right)
+ (\Delta\theta)^2 \left( {\partial A_k \over \partial \theta} \right)
\left( {\partial A_l \over \partial \theta} \right)
\eea

\beq
n_i = \delta n_{LL} , \delta n_{RR} , \delta n_{LR} , \delta n_{RL}
~,
\eeq

where the statistical errors 
originating from the uncorrelated polarized event rates
in each bin are given by

\beq
\label{stat}
\Delta n_i = \sqrt{ n_i}
~,
\eeq
whereas the systematic error (second term in Eq.(\ref{V}))
stems  from the inaccurate measurement
of the scattering angle. A realistic value that we employ in our analysis is

\beq
\label{syst}
\Delta\theta = 0.5 \mbox{ mrad}
~.
\eeq

Since the small angle singularities 
of the differential \xs s
cancel out
in the asymmetries,
the latter have a rather smooth angular dependence.
As a result,
the contribution of the second term in Eq.~(\ref{V})
is almost negligible.

The quadratic form (\ref{egg})
defines a 3-dimensional ellipsoid
in the 
$(\swt, P_1, P_2)$
parameter space.
The inverse square root of the diagonal elements 
of the inverse covariance matrix $W^{-1}$
are the values of the intersections 
of the error ellipsoid with the corresponding parameter axes.
These correspond to the one-standard-deviation errors
on this parameter,
assuming the other two parameters are known exactly.
In contrast,
the square roots of the diagonal elements 
of the covariance matrix $W$
are the values of the projections 
of this ellipsoid onto the corresponding parameter axes.
These correspond to  the one-standard-deviation errors
on this parameter,
whatever values the other two parameters assume.
When presenting our results
we choose the latter for our predictions
of the errors on \swt\ and the beam polarizations.

Unless stated otherwise, 
we assume from now on the following values
for the expectation values of the parameters
and the angular acceptance of the detector:
\beq
\label{numbers}
\left\{
\barr{l}
  \swt = .2315 \\
  P_1=P_2 = 90 \% \\
  |\cos\theta| < .995 
\earr
\right.
\eeq

To take into account the angular dependence of the asymmetries,
we have chosen to work with 200 equal size bins in $\cos\theta$
over the angular range (\ref{numbers}).
This is easy to implement experimentally,
as the scattering angles
can be measured with very high accuracy (\ref{syst}).
Since the asymmetries have a relatively smooth angular behaviour,
increasing the number of bins beyond 50
does not significantly improve the accuracy of the measurement.
We have checked that,
as expected,
the results approach very closely 
the Cram\'er-Rao minimum variance bound \cite{cr}.

\begin{figure}
\hspace*{-3em}
\input{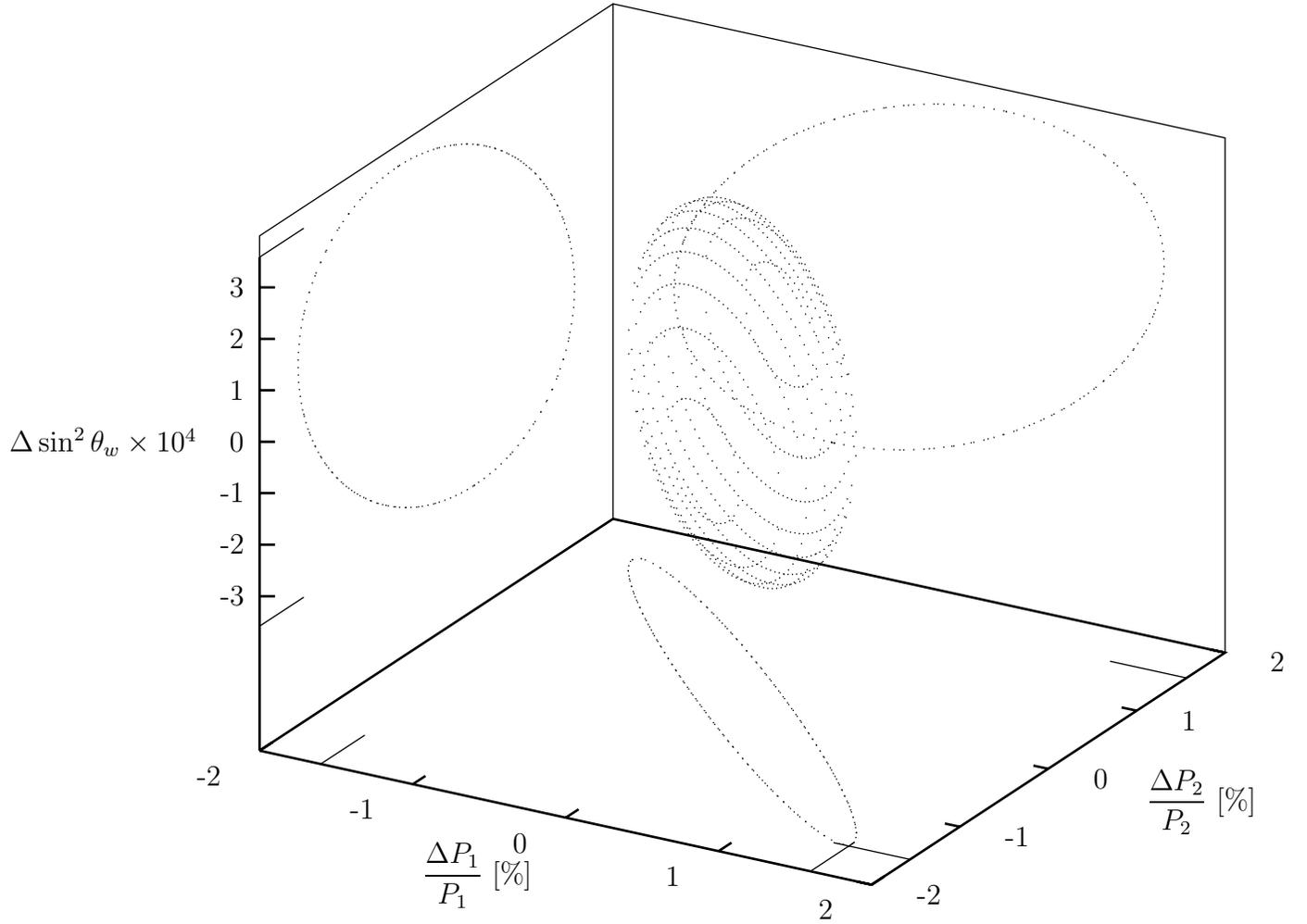}
\caption{
One standard deviation error bounds
on the measurement of 
\swt\ and the beam polarizations
for Bhabha scattering at 500 GeV \cm\ energy.
}
\label{flat}
\end{figure}

For the sake of illustration,
we have plotted in Fig.~\ref{flat}
the 3-dimensional ellipsoid defined by the quadratic form (\ref{egg})
for the \pe\ experiment at 500 GeV.
This figure provides some interesting insight.
For instance,
it is clear that the two polarization measurements
are highly correlated,
in the sense that the average polarization 
can be determined much more precisely
than the polarization difference of the two beams.
In contrast,
\swt\ is only weakly correlated to the beam polarizations.

\begin{figure}
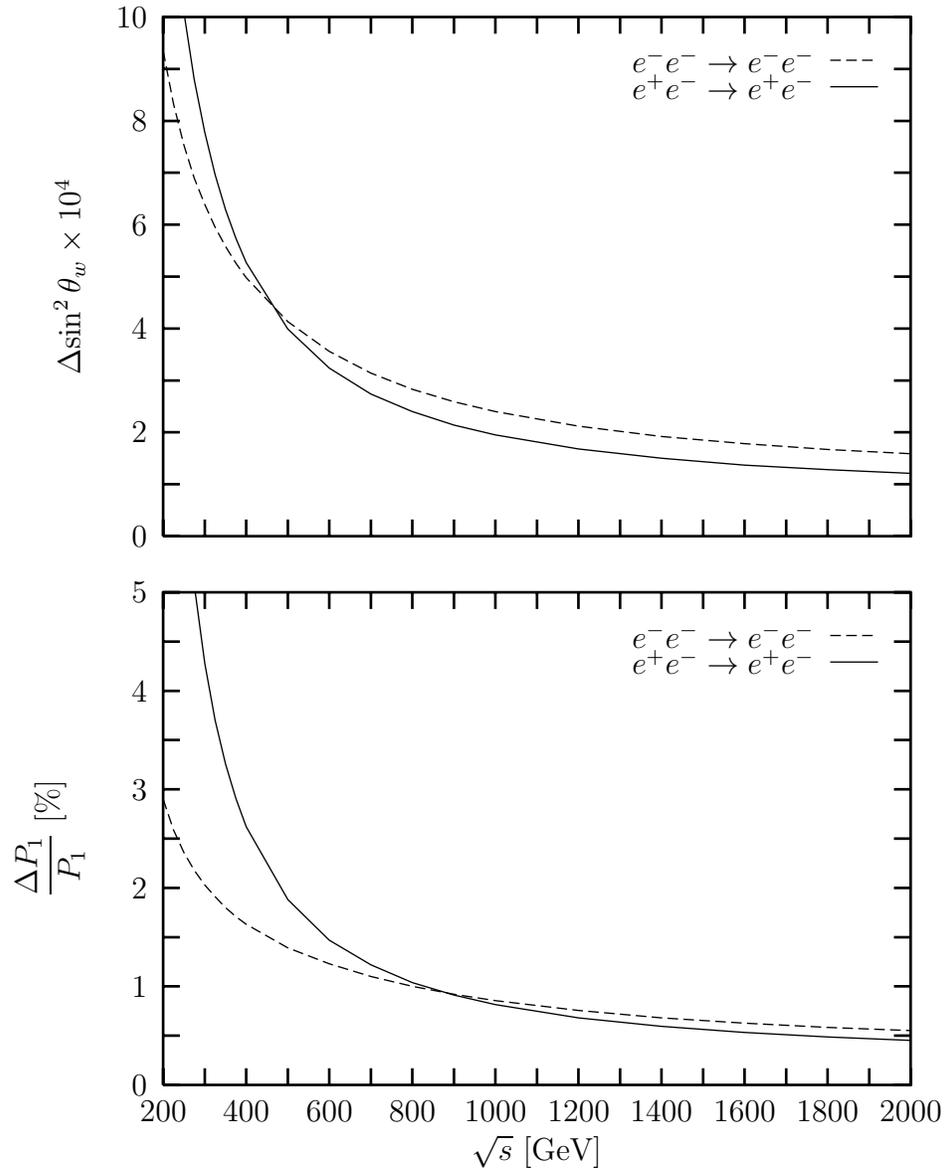

\input{eny.sw2.pstex}
\input{eny.pol.pstex}
\caption{
Energy dependence of the errors on
\swt\ and the beam polarizations
in M\o ller and Bhabha scattering.}
\label{feny}
\end{figure}

Because we assume the luminosities
to scale proportionally to the square of the collider energy 
(\ref{lumpe},\ref{lumee}),
the resolution of the measurement
improves at higher energies.
This is displayed in Fig.~\ref{feny},
where we plot
the \cm\ energy dependence
of the one standard deviation errors
on the measurements of \swt\ and the electron beam polarization.
We observe a clear saturation beyond 1 TeV
for both Bhabha and M\o ller scattering.

At $\sqrt{s}=500$ GeV,
\swt\ can be measured with an error of about $4 \times 10^{-4}$.
Although this will not improve  the combined LEP-SLC accuracy,
it may provide an independent check.
On the other hand,
at 2 TeV the resolution on \swt\ can reach up to $1 \times 10^{-4}$.
Similarly,
the polarization can be determined at 500 GeV
down to 2\%\ in Bhabha
and 1.5\%\ in M\o ller scattering.
Compton polarimetry currently 
yields a similar accuracy 
of 1.7\%\ \cite{slcpol}
and is constantly improving.
However, at 2 TeV
both Bhabha and M\o ller scattering
can measure the polarization down to 0.5\%, 
a very promising result.

\begin{figure}
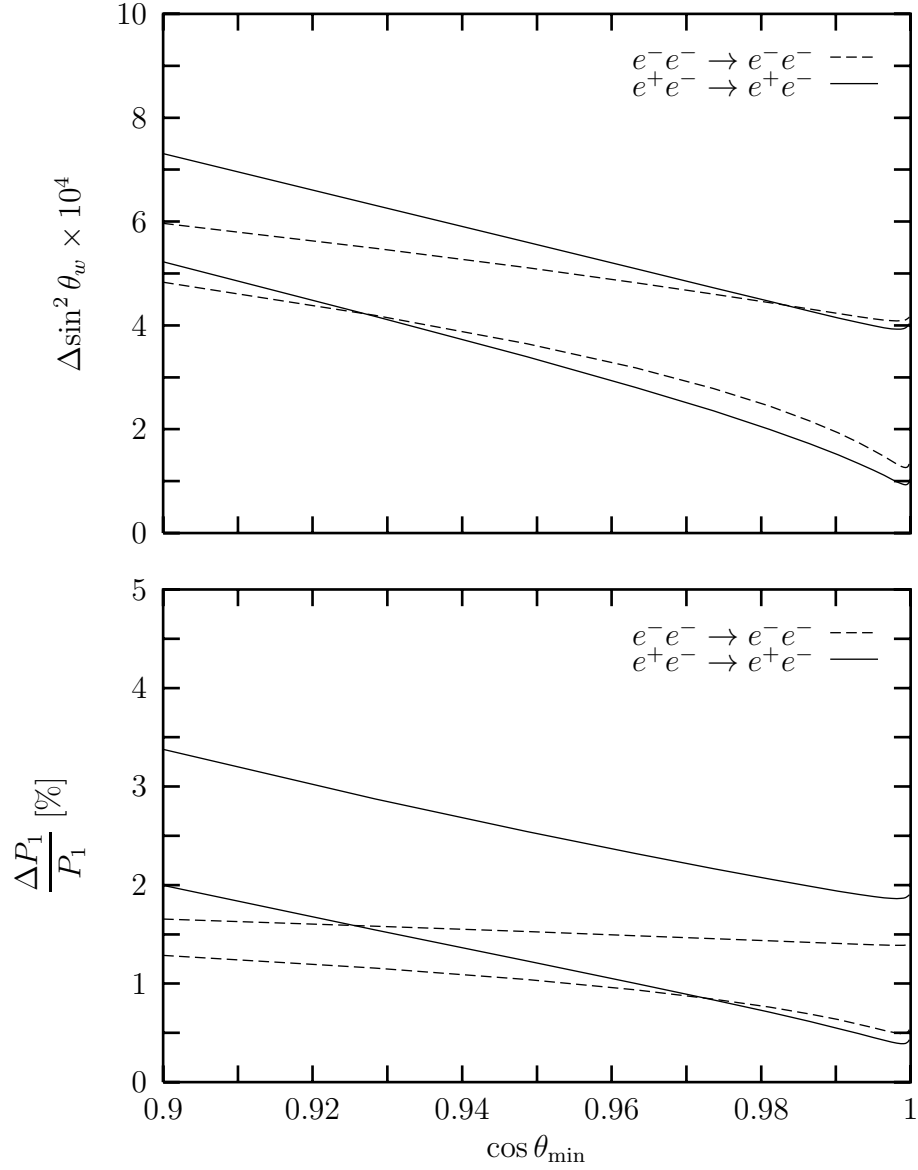

\input{cos.sw2.pstex}
\input{cos.pol.pstex}
\caption{
Polar angle acceptance
dependence of the errors on
\swt\ and the beam polarizations
in M\o ller and Bhabha scattering.
The upper and lower pairs of curves 
correspond to 500 GeV and 2 TeV
\cm\ energy collisions.}
\label{fcos}
\end{figure}

As we mainly rely on the $\gamma-Z^0$ interferences
to measure \swt,
it is essential to probe small scattering angles.
This is well depicted in Fig.~\ref{fcos},
where we display
the errors
as a function of the polar angle coverage.
The slight decrease in sensitivity
observed for very small polar angle
is due to the finite bin size.
Improving the angular coverage 
beyond $5^o$ does not appear  to be very  useful.
The error on the polarization
is not very  sensitive to the detector acceptance,
especially for M\o ller scattering.

\begin{figure}
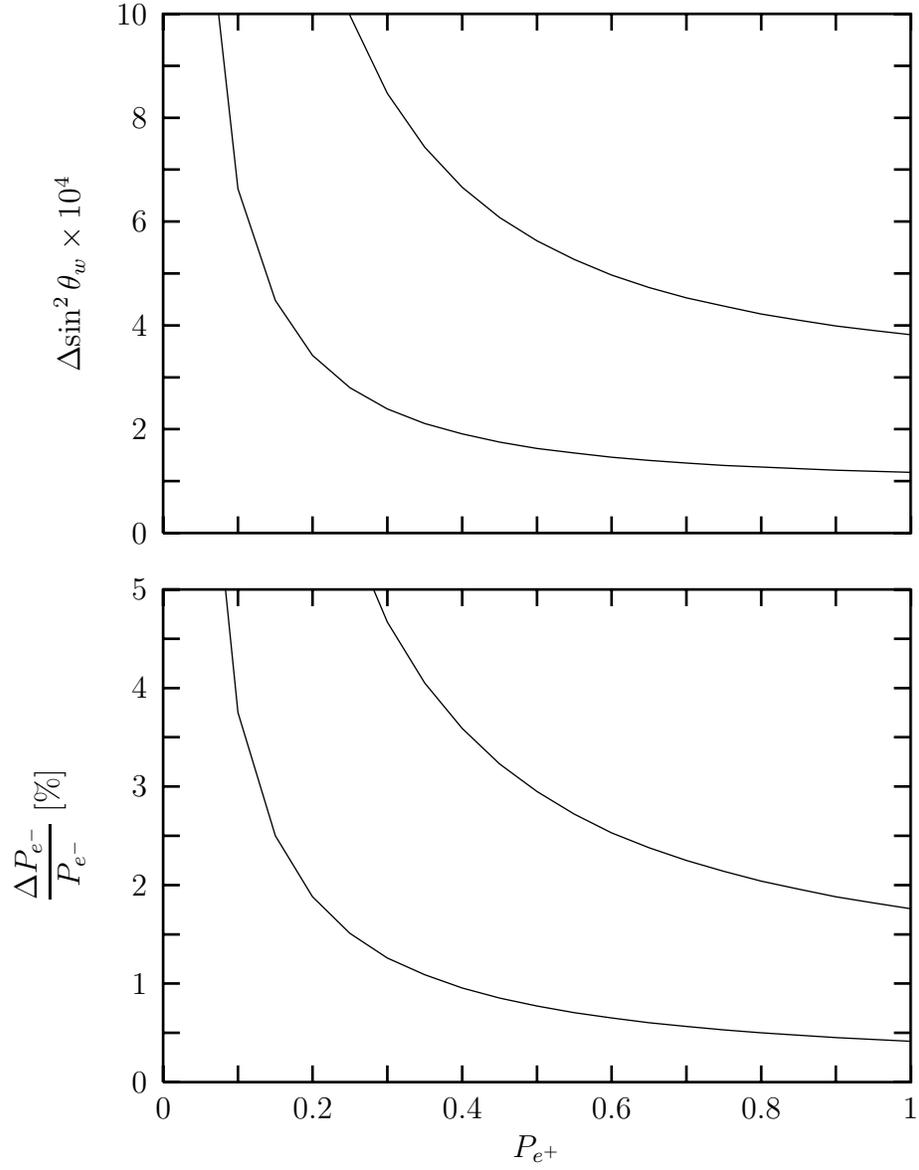

\input{ppo.sw2.pstex}
\input{ppo.pol.pstex}
\caption{
Positron polarization 
dependence of the errors on
\swt\ and the electron beam polarization
in Bhabha scattering.
The upper and lower curves 
correspond to 500 GeV and 2 TeV
\cm\ energy collisions.
}
\label{fppo}
\end{figure}

The dependence of 
the errors
on the polarization of both beams
is displayed in Fig.~\ref{fpol}.
Clearly,
high degrees of polarization are an important asset,
especially at lower energies.
This should not present any problem for the electron beams
and the M\o ller scattering experiment.
To gauge,
however, 
the importance of the positron polarization
in Bhabha scattering,
we plot in Fig.~\ref{fppo}
the errors
as a function of the positron beam polarization.
It appears that
at 500 GeV
the resolution degrades significantly
for positron polarizations
less than 50\%.
For 2 TeV collisions
positron polarizations
as small as 30\%\
still yield interesting results.

\begin{figure}
\input{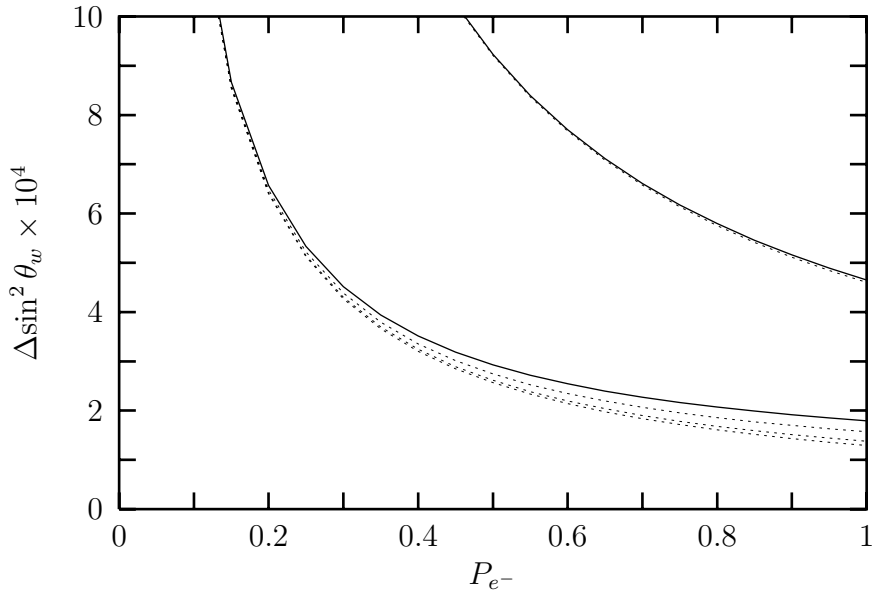}
\caption{
Electron polarization 
dependence of the errors on
\swt\
in Bhabha scattering.
The positron beam is unpolarized
and the polarization of the electrons 
is assumed to have been determined 
with a precision of 1.5\%\
by other means.
The upper and lower curves 
correspond to 500 GeV and 2 TeV
\cm\ energy collisions.
The dotted curves indicate the expectations 
with 1\%, 0.5\%\ and no  error on the electron polarization.
They can  almost not be resolved 
on this scale 
from the 500 GeV curve.
}
\label{fnop}
\end{figure}

In the event the positrons cannot be polarized at all, 
a strong correlation develops
between \swt\ and the electron polarization
so that these two parameters remain effectively unconstrained.
Still,
\swt\ can 
be determined  accurately
if the electron polarization 
is also known precisely from the onset
(from Compton polarimetry for instance)
and its resolution is treated as a systematic error.
In this case 
we observe in Fig.~\ref{fnop},
that the resolution on \swt\ 
is approximately degraded by as little as a factor $\sqrt{2}$.
The systematic error stemming from the measurement
of the electron polarization
is not very important.

The bounds to be obtained 
for a few realistic energies and polarizations
are summarized in Table~\ref{t1}.
They assume of course the validity 
of the luminosities stated in Eqs~(\ref{lumpe},\ref{lumee}).
For different values of the integrated luminosity
the results can be easily  corrected,
since the statistical errors largely dominate the systematic errors
included here
and scale like
$1/\sqrt{\cal L}$. 
We also note that, with respect to the processes studied 
here, the \pe $\rightarrow \mu^+ \mu^-$ mode 
yields much less interesting bounds, 
about one order of magnitude  worse. This 
is obviously expected because of the absence of the forward Coulomb peak
in this case.

The present preliminary analysis has been carried out at the tree level only.
Electroweak radiative corrections to Bhabha scattering off the 
$Z^0$ peak have been first calculated 
in \cite{consoli}, and  updated  to leading two-loop order in \cite{bhr}. 
In general,  electroweak corrections can be included in the Bhabha amplitudes 
by means of three complex-valued gauge invariant 
 form factors explicitly depending on $\theta$ 
\cite{bhr}. We expect a similar factorization of radiative corrections 
in M\o ller scattering, for which a full one-loop computation, 
to the best of our knowledge, is  still missing at high energies
\cite{czarnecki}.
The inclusion of 
these calculable radiative effects and a discussion of the problems connected 
to matching the required accuracy of less than  one permille
are  beyond the scope of this paper, but
they  should not affect  significantly
our estimates of the statistical error, particularly because they are
 dominated by events in the forward peak, where electroweak corrections become 
less relevant.
QED  effects are generally quite sizable in large angle Bhabha and M\o ller
scattering
\cite{QED}, and could in principle introduce additional
 uncertainties. However, soft photons and other QED effects factorize and 
cancel in the parity violating asymmetries, and we
do not expect dramatic effects on our error estimates.

To conclude,
we have demonstrated how the large Bhabha and M\o ller scattering \xs s
can be advantageously used 
at a high energy \lc\
to measure the polarization of the incoming
electron or positron beams
down to the percent level or better.
The method we propose
measures the polarization of the interacting beams
through the final states, 
so that it takes into account all depolarizing effects 
due to beamstrahlung and disruption.

Simultaneously,
the value of \swt\ can be determined
in 500 GeV collisions
with an absolute error of about $4 \times 10^{-4}$.
This error can be further reduced
down to almost $1 \times 10^{-4}$,
by increasing the \cm\ energy up to 2 TeV.
Beyond this energy, however,
there is little gain 
unless the  luminosity is increased with respect to 
Eq.~(\ref{lumpe},\ref{lumee}).

These precision measurements can be easily carried out
and do not interfere with the main tasks of the \lc.
To reach the abovementioned accuracies,
though,
it is essential to have
a good polar angle coverage 
of the detector 
as well as highly polarized beams.

If electron and positron beams
can be polarized with the same efficiency,
both Bhabha and M\o ller scattering
yield very similar results.
At high energies
Bhabha scattering 
performs marginally better,
because of the higher luminosity
of the \pe\ mode
with respect to the \ee\ mode (\ref{lumee}).
However,
if positron beams cannot be polarized, 
the resolving power of Bhabha scattering
is approximately reduced by 30\%.

\section*{Acknowledgments}
We are very much indebted to Peter Zerwas and Ron Settles
for their suggestions and
for pointing out Ref.~\cite{flott} to us. 
We are also grateful to Abdel Djouadi and Ron Settles
for their careful reading of the manuscript.
P.G. would like to thank Sacha Davidson,
Andrea Galli and Bernd Kniehl for interesting discussions.

\clearpage

\newcommand{\sd}[1]{\raisebox{-2.5ex}[0ex][0ex]{$#1$}}
\begin{table} 
$$
\tabcolsep1em
\barr{||c|c|c|c|c|c|c||}
\hline\hline
\mbox{reaction} 
& \sqrt{s} \mbox{ [TeV]} 
& \quad P_1 \quad 
& \quad P_2 \quad 
& \Delta\swt \times 10^4 
& \Delta P_1/P_1 \mbox{ [\%]} 
& \Delta P_2/P_2 \mbox{ [\%]} 
\\\hline\hline
& & 
0 & .9 & 5.2 & - & 1.5^*
\\
\sd{\pe\to\pe} & \sd{.5} &
.3 & .9 & 8.5 & 4.5 & 4.7
\\
& & 
.6 & .9 & 5.0 & 2.5 & 2.5
\\
& & 
.9 & .9 & 4.0 & 1.9 & 1.9
\\\hline
\ee\to\ee & .5 &
.9 & .9 & 4.1 & 1.4 & 1.4
\\\hline
& & 
0 & .9 & 1.9 & - & 1.5^*
\\
\sd{\pe\to\pe} & \sd{2} &
.3 & .9 & 2.4 & 1.2 & 1.3
\\
& & 
.6 & .9 & 1.5 & .65 & .65
\\
& & 
.9 & .9 & 1.2 & .45 & .45
\\\hline
\ee\to\ee & 2 &
.9 & .9 & 1.6 & .55 & .55
\\\hline\hline
\earr
$$
\caption{
One standard deviation error bounds 
on the measurements of \swt\ and the beam polarizations 
in Bhabha and M\o ller scattering 
for various values of energy and polarization.
In the case of Bhabha scattering 
$P_1$ stands for the positron polarization.
When the positrons are not polarized,
the polarization of the electrons 
is assumed to have been determined 
with a precision of 1.5\%\
by other means ($^*$).
}
\label{t1}
\end{table}

\clearpage

\begin{figure}
\input{pol.sw2.pstex}
\input{pol.pol.pstex}
\caption{
Polarization 
dependence of the errors on
\swt\ and the beam polarizations
in M\o ller and Bhabha scattering.
The upper and lower pairs of curves 
correspond to 500 GeV and 2 TeV
\cm\ energy collisions.}
\label{fpol}
\end{figure}

\end{document}